# Section 230: A Juridical History

Gregory M. Dickinson[*]

ABSTRACT

*Section 230 of the Communications Decency Act of 1996 is the most important law in the history of the internet. It is also one of the most flawed. Under Section 230, online entities are absolutely immune from lawsuits related to content authored by third parties. The law has been essential to the internet's development over the last twenty years, but it has not kept pace with the times and is now a source of deep consternation to courts and legislatures. Lawmakers and legal scholars from across the political spectrum praise the law for what it has done, while criticizing its protection of bad-actor websites and obstruction of internet law reform.*

*Absent from the fray, however, has been the Supreme Court, which has never issued a decision interpreting Section 230. That is poised to change, as the Court now appears determined to peel back decades of lower court case law and interpret the statute afresh to account for the tremendous technological advances of the last two decades. Rather than offer a proposal for reform, of which there are plenty, this Article acts as a guidebook to reformers by examining how we got to where we are today. It identifies those interpretive steps and missteps by which courts constructed an immunity doctrine insufficiently resilient against technological change, with the aim of aiding lawmakers and scholars in crafting an immunity doctrine better situated to accommodate future innovation.*

---

[*]Assistant Professor of Law and, by courtesy, Computer Science, the University of Nebraska; Nonresidential Fellow, Stanford Law School, Program in Law, Science & Technology; J.D., Harvard Law School. Thanks to participants at the University of Nebraska College of Law Works in Progress Workshop in October 2024 and to Kate Taylor for her research assistance.

TABLE OF CONTENTS

I. INTRODUCTION

Section 230 of the Communications Decency Act (CDA)[1] has been called the law that created the internet;[2] and it very nearly deserves it. Under Section 230, online platforms, internet service providers, websites, smartphone apps, and other online entities are immune from legal liability for content created by third parties. To see the law's importance, just imagine what the world might look like if the law were otherwise: Who would dare run a platform like Facebook, an online discussion forum like Reddit, or even a website with a comments section if she were at risk of being haled into court any time some bozo on the internet used the system to say something unlawful? The answer might very well be no one. By providing that only creators of online content and not mere conduits can be held legally responsible, Section 230 has guaranteed the ability of countless smartphone apps and websites to flourish without fear of legal liability for material created by others.

That, at least, is one side of the story. By ensuring online entities are not liable for content created by others, Section 230 paved the way for the bustling virtual world that we know today. But the statute has become the victim of its own (and the internet's) success. Section 230 was designed for the publication-centric internet of 1996. It speaks in terms of "publishers," "speakers," and "content providers" and is well-suited to govern the internet's print-like information repositories and communications channels, such as online newspapers, magazines, video streaming services, and the web hosts and online portals that support them. Such entities are direct descendants of the print-media authors, publishers, broadcasters, and bookstores that preceded them and are readily governed by Section 230's publication-centric framework.

Section 230's simple rule for vicarious liability works well in the publication context: Content creators are subject to liability, whereas mere conduits of content are not.[3] But the internet has evolved dramatically from its publication-centric roots. It now includes all manner of virtual-world entities, across the full spectrum of human activity. Often, they do not fall neatly into the creator or conduit bin. When Grindr or Snapchat designs a smartphone app that teenagers predictably misuse to arrange rendezvous with sexual predators[4] or to take selfies while driving recklessly,[5] is the company acting as a content creator, subject to liability, or as a conduit for its users and therefore immune? Realistically it is neither, but Section 230 has no third option. What about Amazon's hosting of product listings by third-party sellers—conduit for others' listings, or creator of its own virtual storefront and subject to suit for defective products?[6] Or telehealth provider hims.com, offering customers pills for "great sex," to "regrow hair," "tackle anxiety," and "lose weight," all eventually prescribed by third-party physicians. Is the company offering medical services and subject to malpractice claims, or is it merely a conduit for third-party physicians and therefore immune?[7] Social media might seem like simple content conveyance. But what of the algorithms that select and organize that content?[8] In short, today's internet entities often defy any simple classification either as content creators or as conduits, yet they enjoy broad immunity under a statute from a different era.

---

[1] Communications Decency Act of 1996, Pub. L. 104-104, 110 Stat. 56, tit. V (1996) (relevant provisions codified at 47 U.S.C. § 230). This Article draws from and synthesizes much of my previous work on Section 230 internet immunity doctrine. *See generally* Gregory M. Dickinson, *The Internet Immunity Escape Hatch*, 47 BYU L. REV. 1435 (2022); *Toward Textual Internet Immunity*, 33 STAN. L. & POL'Y REV. ONLINE 1 (2022); *Rebooting Internet Immunity*, 89 GEO. WASH. L. REV. 347 (2021); Note, *An Interpretive Framework for Narrower Immunity Under Section 230 of the Communications Decency Act*, 33 HARV. J.L. & PUB. POL'Y 863 (2010).
[2] JEFF KOSSEFF, THE TWENTY-SIX WORDS THAT CREATED THE INTERNET (2019).
[3] 47 U.S.C. § 230(c)(1).

---

[4] *See generally* Doll v. Richard Pelphrey & Grindr Inc., No. 23 CI 00363, 2024 Ky. Cir. LEXIS 84 (Ky. Cir. Ct. Oct. 18, 2024) (dismissing negligent product design claim against Grindr as barred by Section 230).

[5] Lemmon v. Snap, Inc., 995 F.3d 1085, 1090–91 (9th Cir. 2021) (concluding Section 230 did not bar negligent product design claim against Snap); Maynard v. Snapchat, Inc., 816 S.E.2d 77, 79–82 (Ga. Ct. App. 2018) (same).

[6] *Compare* Bolger v. Amazon.com, LLC, 267 Cal. Rptr. 3d 601, 626–27 (Cal. App. 2020) (finding Amazon subject to strict liability claims for third-party listings and not immune under Section 230) *and* Oberdorf v. Amazon.com Inc., 930 F.3d 136 (3d Cir.) (same), *reh'g en banc granted*, *opinion vacated*, 936 F.3d 182 (3d Cir. 2019), *question certified*, 818 F. App'x 138 (3d Cir. 2021), *certified question accepted*, 237 A.3d 394 (Pa. 2020), *with* Eberhart v. Amazon.com, Inc., 325 F. Supp. 3d 393, 400 (S.D.N.Y. 2018) (finding Amazon not subject to strict liability claims for third-party listings and immune under Section 230).

[7] *Cf.* Indictment at 11–12, United States v. He, No. 3:24-CR-329 (N.D. Cal. filed Jun. 12, 2024) (alleging defendants' companies maintained a network of medical professionals whom they paid to write prescriptions for controlled substances including Adderall and other stimulants).

[8] *Compare* In re Soc. Media Adolescent Addiction/Pers. Inj. Prod. Liab. Litig., __ F. Supp. 3d __, No. 23-CV-05448, 2024 WL 4532937, at *61 (N.D. Cal. Oct. 15, 2024) (narrowing design defect claims based on Section 230 defense), *with* Anderson v. TikTok, Inc., 116 F.4th 180, 183–84 (3d Cir. 2024) (algorithmic content-selection decisions constitute first-party speech).

The problem has spurred a raft of academic criticism on topics ranging from cyberbullying,[9] online governance,[10] inequality,[11] and freedom of expression,[12] to big-tech antitrust concerns,[13] privacy,[14] and tort liability.[15] Indeed the need for internet immunity reform is an area of rare bipartisan agreement, with reform efforts being led by Democrats, Republicans, and internet law scholars of all stripes.

---

[9] *See* Erica Goldberg, *Free Speech Consequentialism*, 116 COLUM. L. REV. 687, 744–45 (2016) (noting that current internet immunity doctrine bars claims against online entities for revenge porn and other forms of cyberbullying); Andrew Gilden, *Cyberbullying and the Innocence Narrative*, 48 HARV. C.R.-C.L. L. REV. 357, 389–90 (2013) (critiquing proposals to narrow online immunity to protect gay teens from harassment on ground that such efforts obscure the power of individual agency).

[10] *See* Jack M. Balkin, *Free Speech in the Algorithmic Society: Big Data, Private Governance, and New School Speech Regulation*, 51 U.C. DAVIS L. REV. 1149, 1182–93 (2018) (discussing platforms' role as regulators of free speech in digital era); Jennifer Daskal, *Speech Across Borders*, 105 VA. L. REV. 1605, 1637–44 (2019) (discussing geographic scope of online platforms' content-filtering determinations and implications for territorial sovereignty); Kate Klonick, *The New Governors: The People, Rules, and Processes Governing Online Speech*, 131 HARV. L. REV. 1598, 1599–613 (2018) (tracing the ability of private platforms like Facebook to make content-moderation decisions regarding user-submitted content to Section 230); Frank Pasquale, *Two Narratives of Platform Capitalism*, 35 YALE L. & POL'Y REV. 309, 316–19 (2016) (offering two possible narratives of the distributed online platform and implications for each on regulatory and self-governance policy decisions); *see also* David R. Johnson & David Post, *Law and Borders—The Rise of Law in Cyberspace*, 48 STAN. L. REV. 1367, 1367 (1996) (arguing just after Section 230's enactment that internet regulation would require its own distinct principles); Lawrence Lessig, *The Law of the Horse: What Cyberlaw Might Teach*, 113 HARV. L. REV. 501, 502 (1999) (arguing that the study of cyberlaw can illuminate principles that affect the real world).

[11] Olivier Sylvain, *Platform Realism, Informational Inequality, and Section 230 Reform*, 131 YALE L.J. FORUM 475 (2021) (discussing disproportionate effects of Section 230 immunity on historically marginalized groups).

[12] *See* Danielle Keats Citron, *Extremist Speech, Compelled Conformity, and Censorship Creep*, 93 NOTRE DAME L. REV. 1035, 1036–40 (2018) (exploring the departure of online platforms from U.S. First Amendment values and the dangers of bowing to international pressure to self-regulate); Danielle Keats Citron & Helen Norton, *Intermediaries and Hate Speech: Fostering Digital Citizenship for Our Information Age*, 91 B.U. L. REV. 1435, 1453–84 (2011) (noting that Section 230 insulates platforms from legal liability and offering proposals for online platforms to voluntarily respond to online hate speech); Eric Goldman, *Why Section 230 Is Better Than the First Amendment*, 95 NOTRE DAME L. REV. REFLECTION 33, 36–46 (2019) (discussing Section 230's enhanced substantive and procedural protections for online entities beyond those of the First Amendment); Rebecca Tushnet, *Power Without Responsibility: Intermediaries and the First Amendment*, 76 GEO. WASH. L. REV. 986, 1009 (2008) (arguing that Section 230 immunity should include a corresponding limit on an intermediary's ability to censor speech); Felix T. Wu, *Collateral Censorship and the Limits of Intermediary Immunity*, 87 NOTRE DAME L. REV. 293, 295–96 (2011) (noting speech-enhancing effects of Section 230 due to its preventing imposition of liability on intermediaries for harmful or offensive speech that those intermediaries might otherwise be pressured to censor).

Rather than add to the large and growing body of work (my own included) discussing how Section 230 should be reformed, this Article steps back to address a more fundamental question: How did a statute designed to limit defamation and other publication-related claims against online service providers come to protect websites and apps against claims as varied as products liability, sex trafficking, aiding and abetting of terrorism, and antitrust violations?

---

[13] *See* C. Scott Hemphill, *Disruptive Incumbents: Platform Competition in an Age of Machine Learning*, 119 COLUM. L. REV. 1973, 1974–93 (2019) (identifying potential sources of competition among dominant participants in online platform market and offering proposals to maximize competition); Lina M. Khan, *The Separation of Platforms and Commerce*, 119 COLUM. L. REV. 973, 1037–92 (2019) (proposing bars on entities' engaging in new lines of business as a check on dominance of a small number of tech firms); Lina M. Khan & David E. Pozen, *A Skeptical View of Information Fiduciaries*, 133 HARV. L. REV. 497, 527–28 (2019) (noting Google and Facebook's capture of the digital advertising market in the United States and resultant effects on the traditional publishing industry).

[14] *See* Danielle Keats Citron, *How to Fix Section 230*, 103 B.U. L. REV. 713 (2023) (proposing modification to Section 230 immunity to spur platforms to action to protect against cyber stalking and other intimate privacy violations); Danielle Keats Citron, *Sexual Privacy*, 128 YALE L.J. 1870, 1952–53 (2019) (revenge porn and other invasions of sexual privacy); Bobby Chesney & Danielle Citron, *Deep Fakes: A Looming Challenge for Privacy, Democracy, and National Security*, 107 CALIF. L. REV. 1753, 1755–59, 1795–804 (2019) (describing rising danger to privacy and security posed by advances in technology for creating deep fakes and noting that Section 230 limits legal recourse against online entities that distribute such fakes).

[15] *See* Ann Bartow, *Internet Defamation as Profit Center: The Monetization of Online Harassment*, 32 HARV. J.L. & GENDER 383, 384 (2009) (tracing the rise of commercial reputation defense services to the lack of traditional avenues of recourse to respond to online harassment); Danielle Keats Citron, *Mainstreaming Privacy Torts*, 98 CALIF. L. REV. 1805, 1836–43 (2010) (sketching the vision for a new era of privacy law and noting the barrier that Section 230 poses to tortious enablement claims against online entities); Danielle Keats Citron & Benjamin Wittes, *The Problem Isn't Just Backpage: Revising Section 230 Immunity*, 2 GEO. L. TECH. REV. 453, 455–56 (2018) (proposing that online immunity be narrowed to allow claims against online entities that do not take reasonable steps to address unlawful third-party content); Benjamin Edelman & Abbey Stemler, *From the Digital to the Physical: Federal Limitations on Regulating Online Marketplaces*, 56 HARV. J. ON LEGIS. 141, 143 (2019) (noting the bar that Section 230 poses to the regulation of modern online marketplaces); Olivier Sylvain, *Intermediary Design Duties*, 50 CONN. L. REV. 203, 203 (2018) (suggesting that the online immunity doctrine be updated to consider the manner in which online entities elicit and use their users' data).

The Article shows that courts' current, broad interpretation of Section 230 was far from inevitable. Instead, it proceeded in a series of three discrete interpretive steps: First, courts interpreted Section 230 to operate independently of the Communications Decency Act's primary goal of limiting children's access to indecent material, and to apply regardless of an entities' effort to filter offensive material; second, they interpreted the statute to depart from the common law and to immunize entities even against claims of their knowing, rather than merely unintentional, distribution of unlawful content; and third, they interpreted the statute to grant immunity against nearly any cause of action, not only those, like defamation, with a historical connection to the publication industry.

At each stage of evolution, courts made textually defensible, but questionable, choices to broaden the scope of Section 230 immunity, largely on the ground that doing so would help then-new internet technologies to thrive unimpeded by the prospect of legal liability. In nurturing a nascent internet, courts succeeded. Unfortunately, their key interpretive moves were made in the 1990s and 2000s and without sufficient nuance to account for the changes that would soon be brought by smartphones, AI, and the gig economy.

Although courts are beginning to recognize the error, the case law has by now become ossified, and the Supreme Court appears determined to step in. Its case selection and comments in dicta over the last four years reveal a Court prepared for the first time since the 1990s to peel back the case law to the statute's core and interpret the text afresh, now with decades of hindsight.[16] This Article serves as a reference point for that undertaking, analyzing the interpretive moves that brought us to this point and how the law might have been and still could be otherwise.

Part II situates Section 230 in its historical context by tracing the evolution of defamation law in the twentieth century to account for the rise of radio, television, and other mass media, which developments were the direct inspiration for the 1996 statute; Part III analyzes the three key interpretive moves by courts in the 1990s and 2000s, which transformed Section 230 from a narrowly focused statute into the comprehensive internet immunity provision it is today; and Part IV concludes by identifying those cases most problematic to current Section 230 doctrine and sketching potential paths forward for reform.

---

[16] *See* Gonzalez v. Google LLC, 598 U.S. 617, 622 (2023) (granting certiorari to resolve question regarding Section 230 but deciding the appeal on other grounds); Twitter, Inc. v. Taamneh, 598 U.S. 471, 507 (2023) (same); *see also* Malwarebytes, Inc. v. Enigma Software Grp. USA, 141 S. Ct. 13, 13–18 (2020) (statement of Thomas, J. regarding denial of certiorari); Biden v. Knight First Amend. Inst. At Columbia Univ., 141 S. Ct. 1220, 1221–27 (2021) (Thomas, J., concurring).

## II.    THE PRE-INTERNET COMMON LAW

The internet and the apps and websites it powers may be new, but the problems they raise span millennia. To understand Congress's organization of Section 230 and early courts' interpretations of its key provisions in the 1990s, it is necessary first to go back even further, to consider tort law's evolution in the twentieth century, especially the law of defamation, to accommodate the distance-spanning technologies that drive mass media.

### A.    Defamation and Republication

Anyone who has felt the sting of a false accusation knows that sharp words can cause just as much hurt as a sharp elbow, sometimes more. And being far easier than elbows to transmit over long distances, words are a primary means of causing harm online. Perhaps the most common way that individuals hurt each other online is by making false statements that harm another's personal reputation or the reputation of her business. Both online and offline, defamation law has long protected against such reputational harms. An individual harmed by another's false statement about her can sue to recover damages as long as she can show (1) defendant's publication of defamatory material, (2) about her, (3) to one or more third persons.[17]

Like other torts, defamation law provides recourse to those harmed by others' wrongful conduct. Defamation is unusual, however, in that its harms never come completely to rest. Gossip spreads like wildfire, and a mistaken or vicious comment made in private conversation can easily spread to an entire neighborhood, continually dragging the victim's name through the mud each time the falsehood is uttered afresh. To account for the continuing effects of defamation, the common law treated every new repetition of a defamatory statement as a new act of defamation, subject to the same liability as the first.[18] Not only the original speaker, but also anyone who repeated or republished a defamatory statement would be liable for the harms caused to the victim's reputation, even if the subsequent speaker had no reason to know that the statement was false.[19] This strict republication rule helped to control the spread of reputation-damaging rumors by placing all re-speakers and re-publishers at risk of liability if the original statement turned out to be untrue.

---

[17] *See, e.g.*, Foster v. Churchill, 665 N.E.2d 153, 157 (N.Y. 1996) (defining defamation as "the making of a false statement which tends to expose the plaintiff to public contempt, ridicule, aversion or disgrace"); *see also* RESTATEMENT (SECOND) OF TORTS § 558 (AM. L. INST. 1977); ROBERT D. SACK, SACK ON DEFAMATION § 2:1 (5th ed. 2017).

### B. Mass-Media Conduits

Times have changed significantly from the defamation action's beginnings in the early sixteenth century;[20] the tort has necessarily evolved over time to accommodate new circumstances. One major pressure has been the advent of mass-communication and social-media technologies. A quick look through any social-media newsfeed shows that modern gossip spreads just as fast as always, the juicier the better. Yet the old republication rule designed to quell the gossip wildfires became untenable in the age of mass media.

With the widespread adoption of the telegraph and other mass-communications technologies in the twentieth century, courts soon confronted difficult questions about how to integrate the old rules of defamation law with these promising new technologies. One key question was how to handle the occasional defamatory news story by reputable wire services like the Associated Press (AP). The AP's business model was to author stories on issues of widespread interest and then transmit those stories by telegraph to be reprinted by newspaper publishers around the country.[21] Typically this model worked quite well. Local newspapers were saved the expense of hiring journalists to cover important, but distant stories; and the AP was able to invest greater resources in its newsgathering efforts because it sold its stories to a nationwide audience. But what if the stories end up including defamatory content? A newspaper's publication of an AP-authored story constitutes a repetition of an original speaker's statement, which, under the common law's strict republication rule would make the newspaper liable for the content of all stories it published, even those authored by third-party wire services. Some courts worried that permitting liability against newspaper publishers would slow the growth of new technologies like the telegraph, reasoning that no newspaper could "afford to warrant the absolute authenticity of every item of its news . . . [and still satisfy] the demands of modern necessity for prompt publication."[22]

---

[18] *See* Brent Skorup & Jennifer Huddleston, The Erosion of Publisher Liability in American Law, Section 230, and the Future of Online Curation, 72 OKLA. L. REV. 635, 638–39 (2020).
[19] RESTATEMENT (FIRST) OF TORTS § 580 (AM. L. INST. 1938).
[20] After centuries of resistance to claims predicated on "mere words" (thought more appropriate to the ecclesiastical courts) the English royal courts began to recognize a general cause of action for defamation in the first two decades of the sixteenth century. JOHN BAKER, AN INTRODUCTION TO ENGLISH LEGAL HISTORY 467 (5th ed. 2019). For an early example, see Housden v. Stoyton, LI MS. Maynard 87, fol. 87v; BL MS. Add. 15941, fol. 28 (1568) *reprinted in* JOHN BAKER, BAKER AND MILSOM SOURCES OF ENGLISH LEGAL HISTORY: PRIVATE LAW TO 1750, 698 (2d. ed. 2010) (permitting claim based on defendant's statement that plaintiff "had had the plague in his house and buried in his garden some people who had died of the plague").
[21] Skorup & Huddleston, *supra* note 18, at 639–40.

Although courts initially split on the question, over time a consensus emerged: Defamation law's strict republication rule should not apply to newspaper publications of wire-service news stories.[23] Instead, courts characterized such republications as mere "emanations" of statements whose "true authors" were located elsewhere.[24] Under that principle, which became known as the "wire service defense," the AP could be liable if it authored a defamatory story, but a newspaper who republished the story could be liable only if "it acted in a negligent, reckless, or careless manner in reproducing the story."[25] Defamation law thus joined the general trend of tort law in the early twentieth century, which was transitioning away from its history of strict liability to favor fault-based liability.[26]

During the mid and late twentieth century, the wire service defense evolved in several ways. First, it expanded to include not only wire services but also other mass-media technologies including radio and television.[27] Second, as the term "wire service defense" no longer fit, the principle came to be known instead as "conduit liability" doctrine.[28] The idea was that unlike a traditional publisher, if an entity acted as a mere conduit of another's speech, it would not be held strictly liable and instead would be liable only if it was in some way at fault, for instance, because it knew of the defamatory nature of the material but allowed it to be transmitted anyway. Third, conduit liability was sometimes applied even where an entity was not a pure conduit, merely conveying the speech of some third party, but had also engaged in some editorial discretion, such as censoring select content.[29] Fourth, and finally, during this same period the Supreme Court added a constitutional dimension to conduit liability, recognizing in a series of cases that not only would strict defamation liability impose practical difficulties for publishers, who would struggle to moderate large volumes of content, but that it also would harm the public at large by depriving it of access to protected speech.[30]

---

[22] Layne v. Tri. Co., 146 So. 234, 237 (Fla. 1933) (an early leading case so reasoning and declining to hold a newspaper liable for republishing a defamatory dispatch from a wire service).
[23] Skorup & Huddleston, *supra* note 18, at 639.
[24] Layne, 146 So. at 237.
[25] *Id.*
[26] *See* RICHARD A. EPSTEIN, TORTS §§ 3.2–3.4 (1999) (tracing the common law's shift in the nineteenth and twentieth centuries from strict to fault-based liability for accidental physical invasions to persons and property).

### C. Cubby and Stratton Oakmont

As the dust was finally settling on the old debate over strict liability versus fault-based liability for mass media, a new set of internet-driven publication technologies appeared that would change publication forever. Many early internet companies mimicked their mass media forebears in that their primary function was to convey third-party-authored content to readers.[31] Unlike traditional publishers, however, printing costs can become a nonfactor for online publishers. The low cost of electronic distribution frees online media companies from the role of content gatekeeper and allows them to distribute content at much greater volume. Material need not be screened for quality because electronic distribution costs almost nothing. Companies can, if they wish, publish everything and let the reader decide what to read. Thus, although the new online media companies of the 1990s resembled traditional media in some respects, their effect was to radically democratize publication by making it possible for any person to communicate her ideas, even poorly written, ill-informed, or disfavored ones.[32] As their users shared more and more content over the internet, online companies soon found themselves in court.

---

[27] Skorup & Huddleston, *supra* note 18, at 644–46.
[28] *Id.* at 644.
[29] *Id.* at 644–46.
[30] *Id.* at 646–48; *see also* KOSSEFF, *supra* note 2 at 11–35.
[31] *See* Dickinson, *Rebooting Internet Immunity*, *supra* note 1, at 367–72. (discussing the early internet's evolution from its publication-centric roots).

The first major case was *Cubby v. CompuServe*.[33] At the time, CompuServe was a leading online service provider that offered dial-up access to its network and database of online content. Among that content was the daily newsletter Rumorville, which CompuServe licensed from a third party and made available to its subscribers. The lawsuit arose when Rumorville published allegedly defamatory statements about a competitor, Skuttlebut, calling it a "start-up scam" and claiming that it had stolen Rumorville's stories "through some back door."[34] Skuttlebut's developer, Cubby, sued for defamation, naming as defendants not only the newsletter's author, but also CompuServe, which had made the story available to its subscribers. The court rejected the plaintiffs' claim against CompuServe, reasoning that CompuServe was only a distributor of the newsletter and that it could not be liable unless it "knew or had reason to know of the [defamatory] statements."[35] *Cubby* thus followed pre-internet common law in rejecting strict liability and requiring some level of fault for defamation claims against mass media defendants.

In 1995 a second major case was decided that surprised everyone. In *Stratton Oakmont v. Prodigy Servs.*,[36] a New York state trial court held then-popular online service provider Prodigy liable for defamatory content posted by a third party to one of the service's message boards. The court noted the *Cubby* decision favorably, but distinguished it, reasoning that unlike CompuServe, Prodigy held itself out to the public as a family-friendly, carefully controlled and edited provider and took steps to screen offensive content.[37] Thus, the company had taken on the role of a publisher rather than a mere conduit or distributor and could therefore be strictly liable in a defamation action, even without actual or constructive knowledge of the defamatory content it conveyed. By filtering some objectionable content, the court reasoned, Prodigy had effectively taken ownership of all of it.[38]

---

[32] *See generally* Eugene Volokh, *Cheap Speech and What It Will Do*, 104 Yale L.J. 1805, 1806–07 (1995) (noting that historically the right to free speech has favored popular or well-funded ideas, but predicting that new information technologies would change that).

[33] Cubby, Inc. v. CompuServe, Inc., 776 F. Supp. 135 (S.D.N.Y. 1991).

[34] *Id.* at 138.

[35] *Id.* at 141.

[36] Stratton Oakmont, Inc. v. Prodigy Servs. Co., No. 31063/94, 1995 WL 323710, at *5 (N.Y. Sup. Ct. May 24, 1995).

[37] *Id.* at *4.

The ruling in *Stratton Oakmont* was terrifying to online services because its logic threatened to hold companies liable for all of the content created by their users, who might number in the millions, if the services made efforts to curate that content.[39] If *Stratton Oakmont*'s logic won the day, online companies would have only two options to avoid liability: Spend vast sums of money to review and censor unlawful content, or else take a hands-off approach to ensure they could not be treated as the contents' publishers. Neither option would be good for the internet.

### III. SECTION 230'S ENACTMENT AND EXPANSION

#### A. The Communications Decency Act

In 1995, the internet was expanding rapidly, promising to provide more content to more people than ever before. But with the good came also the bad. Parents in the United States worried that unrestricted access to internet pornography would be harmful to children. Spurred to action to address the internet pornography panic, Congress, led by Senator Exon of Nebraska, started drafting the Communications Decency Act to limit children's access to online pornography by making it unlawful for entities to knowingly make "indecent" material available to minors.[40] Congress at that time was in the midst of implementing a major overhaul to the nation's telecommunications law,[41] so the CDA was added to that package of legislation and became a tiny part of the much larger Telecommunications Act of 1996.[42]

---

[38] *Id.* at *4–5.

[39] *See* KOSSEFF, *supra* note 2, at 53–56.

[40] *See* Communications Decency Act of 1996, Pub. L. No. 104-104, tit. V, 110 Stat. 56, tit. V (1996), declared unconstitutional as to indecent material, Reno v. ACLU, 521 U.S. 844 (1997); *see also* Dickinson, *The Internet Immunity Escape Hatch*, *supra* note 1, at 1453–56; KOSSEFF, *supra* note 2, at 61–66 (describing the social and political backdrop against which the Communications Decency Act of 1995 was enacted).

[41] *See* Ted Stevens, *The Internet and the Telecommunications Act of 1996*, 35 HARV. J. ON LEGIS. 5, 9–18 (1998); *see also* Larry Pressler, *A Look Back at the Telecommunications Act of 1996*, THE HILL, (Feb. 7, 2017), https://perma.cc/JGM5-TKVZ.

[42] Pub. L. No. 104-104, 110 Stat. 56 (1996) (codified primarily in scattered sections of 47 U.S.C.).

The *Stratton Oakmont* decision was issued while work on the CDA was already in progress. When Representatives Chris Cox and Ron Wyden in Congress learned of the decision, they worried about its effect on then-nascent internet technology. Like others in Congress, they recognized internet pornography as problem, and they feared that if *Stratton Oakmont*'s logic gained acceptance in other courts, it could make the problem worse: Imposing liability only against entities that attempted to censor content would create a perverse incentive for companies to avoid moderating content at all.[43] A little over a month after the *Stratton Oakmont* decision, in June 1995, Cox and Wyden proposed the Internet Freedom and Family Empowerment Act,[44] the bill that would eventually become Section 230.

Like Senator Exon's CDA proposal, the bill was intended to reduce children's access to online pornography. But to the CDA's punitive approach, the Cox-Wyden proposal added a market-based component. Section 230 would reduce children's access to pornography by removing a potential disincentive for online companies to censor—the threat of legal liability under *Stratton Oakmont*. The key provisions of the law as enacted read as follows:

> Sec. 230 Protection for Private Blocking and Screening of Offensive Material
>
> . . .
>
> (c) Protection for "Good Samaritan" blocking and screening of offensive material
>
> (1) Treatment of publisher or speaker
>
> No provider or user of an interactive computer service shall be treated as the publisher or speaker of any information provided by another information content provider.
>
> (2) Civil liability

---

[43] *See* Jeff Kosseff, *A User's Guide to Section 230, and A Legislator's Guide to Amending It (or Not)*, 37 BERKELEY TECH. L.J. 761, 767–73 (2022).
[44] H.R. 1978, 104th Cong. (1995).

> No provider or user of an interactive computer service shall be held liable on account of—
>
> (A) any action voluntarily taken in good faith to restrict access to or availability of material that the provider or user considers to be obscene, lewd, lascivious, filthy, excessively violent, harassing, or otherwise objectionable, whether or not such material is constitutionally protected . . .
>
> (f) Definitions
>
> . . .
>
> (3) Information content provider
>
> The term "information content provider" means any person or entity that is responsible, in whole or in part, for the creation or development of information provided through the Internet or any other interactive computer service.[45]

Section 230 thus supported the CDA's major goal, but in a slightly roundabout way. The law promotes "decency" on the internet by allowing online entities to censor "obscene, lewd, lascivious, filthy, excessively violent, harassing, or otherwise objectionable" content without fear of being "treated as the publisher or speaker" of—and held liable for—whatever content they fail to censor.[46] The law protected the nascent internet by guaranteeing online entities' ability to relay and host the massive volumes data flowing through their systems without incurring liability for their content.[47] Absent Section 230's protections, online platforms would face an economically crippling duty to review the inconceivable volume of data that flows through their systems to ensure that none of their users' posts contained defamatory speech or other unlawful content.[48] Online platforms might be compelled to heavily censor user speech or disallow online posting altogether to avoid the risk of liability.

---

[45] 47 U.S.C. § 230(c), (f)(3).
[46] 47 U.S.C. § 230(c).
[47] *See* 47 U.S.C. § 230(a), (c); Dickinson, *Rebooting Internet Immunity*, *supra* note 1, at 360–63 (explaining how the text of Section 230 immunizes internet platforms from tort liability for the content posted by their users).

Despite its current pride of place, however, Section 230's success was something of an accident. The law was tacked on as an afterthought to the Communications Decency Act, which was itself a tiny part of the much larger Telecommunications Act of 1996. And the law underwent little analysis, deliberation, or public debate, for the internet industry of the time was simply too small and the provision too technical to attract significant attention.[49] What would become the most important law in the history of the internet glided through the legislative process virtually unnoticed.

President Clinton signed the act into law on February 8, 1996. Only a year and a half later, the Supreme Court struck down the vast majority of the CDA in *Reno v. American Civil Liberties Union*, reasoning that its criminalization of indecent materials violated the First Amendment.[50] In the end, Section 230 and its key twenty-six words were nearly all that remained from the four-thousand-word CDA.[51]

### B.  Judicial Elaboration of Section 230

Successful as it has been, Section 230 now shows its age. The 1996 statute is designed for an internet that functions to distribute third-party content. Like mass-media distributors before them, online entities often serve as avenues of access to third-party-created content. Section 230 follows tort law's historical treatment of mass-media distributors and grants online entities broad immunity from claims related to third-party content. To do so, Section 230 divides the online world into three camps: (1) "information content provider[s]," analogous to traditional publishers, who create informational content for internet readers; (2) providers of "interactive computer service[s]," such as CompuServe, Prodigy, and, today, Verizon and Comcast, that provide users with the internet access necessary to read that informational content; and (3) users of "interactive computer service[s]," the internet users who consume informational content.[52]

---

[48] *See* Dickinson, *Rebooting Internet Immunity*, *supra* note 1, at 363–64 (explaining that "Section 230 shields online entities from an economically crippling duty to moderate the content flowing through their systems" and that, "[b]y immunizing online entities against lawsuits related to third-party content, Section 230 ensures that the costs of moderating user-created content do not stifle the growth of internet platforms"); *see also* Felix T. Wu, *Collateral Censorship and the Limits of Intermediary Immunity*, 87 NOTRE DAME L. REV. 293, 295–96 (2011) (noting that if online platforms were not given immunity for the unlawful content posted by third parties in their websites, they might engage in collateral censorship and remove even lawful content due to the fear of suit).

[49] *See* KOSSEFF, supra note 2, at 67–71 (explaining that because the internet was new and poorly understood, the House approved the inclusion of Section 230 with little deliberation or media scrutiny).

[50] 520 U.S. 844, 874 (1997) (concluding that "the CDA lacks the precision that the First Amendment requires when a statute regulates the content of speech" because while aiming to protect minors, it "effectively suppresses a large amount of speech that adults have a constitutional right to receive and to address to one another").

[51] 47 U.S.C. § 230(c)(1) ("No provider or user of an interactive computer service shall be treated as the publisher or speaker of any information provided by another information content provider.").

In this neatly divisible, publication-centric world, wrongdoers are easy to identify. Any wrongdoing must be attributable to its active participants—the internet content creators who author content—not the passive internet service providers and their users who merely provide access to and view that content. The solution seemed obvious and uncontroversial when Congress enacted it. Section 230 is a rejection of the *Stratton Oakmont* decision in favor of *Cubby* and the pre-internet distributor and conduit-liability limitations from which it drew.[53] With Section 230 Congress immunized the internet's passive participants from liability by mandating that none but an "information content provider" be held accountable for internet content.[54] But in recent years, Section 230's simple formula has become a recipe for controversy.

Part of the problem is technological change. The internet was never entirely publication centric or so neatly divisible as Section 230 would have it. While Section 230 was being debated in Congress, e-commerce was starting to take root. Amazon.com sold its first book in July 1995,[55] and AuctionWeb, later to become eBay, made its first sale just a few months later, in September 1995.[56] These and other online entities quickly began to strain Section 230's tripartite framework, which is designed for a publication-centric internet comprised of content authors, transmitters, and readers—not for the complete virtual world of smartphones, social media, e-commerce, and online services. Categorizing all internet entities as either content authors, and potentially liable, or as nonauthors, and therefore immune, makes little sense when virtual world wrongdoing often involves no content creation at all.

---

[52] *Id.*
[53] *See supra* Part II.C; Dickinson, *Rebooting Internet Immunity*, *supra* note 1, at 367–72.
[54] 47 U.S.C. § 230(c)(1).
[55] *See generally* RICHARD L. BRANDT, ONE CLICK: JEFF BEZOS AND THE RISE OF AMAZON.COM (2011).
[56] For a history of eBay's early days, see generally ADAM COHEN, THE PERFECT STORE: INSIDE EBAY (2002).

The other part of the problem is Section 230's early treatment in the courts. Section 230 is a "thoroughly ambiguous statute,"[57] amenable to a wide array of interpretations.[58] At one extreme, its language could have been interpreted very narrowly, as a protection merely for those entities that attempt to censor offensive conduct[59] or as a reaffirmation of protections already afforded to distributors and conduits under preexisting law.[60] At the other extreme, the statute can be interpreted broadly, to immunize all online entities against nearly any lawsuit related to third-party-created content. It is the second direction that courts have taken.[61] At every crossroads courts have interpreted the statute's ambiguities to favor broader immunity. The result is an expansive immunity doctrine oddly untethered from the statute that gave it life. Section 230's evolution proceeded in a series of discrete intellectual steps.

### 1. CDA Immunity Without the Decency

First, courts interpreted Section 230's purpose of promoting free expression and internet growth to operate independently of its purpose to promote online decency. [62] Section 230, recall, was added to the "Communications Decency Act" to reduce children's access to online pornography. It was the carrot to Senator Exon's stick. [63] The idea was to supplement the CDA's punitive provisions and encourage voluntary censorship by removing the threat of legal liability for entities, like Prodigy, that sought to offer "family-oriented" online environments.[64] Yet, as it has been interpreted by the courts, Section 230 does almost nothing to achieve that objective.

---

[57] *See generally* Alan Z. Rozenshtein, *Interpreting the Ambiguities of Section 230*, 41 YALE J. REG. BULLETIN 60 (2024) (making this point, especially as to social-media platforms and content-amplification algorithms).

[58] *See* Dickinson, *The Internet Immunity Escape Hatch*, *supra* note 1, at 1461–78 (enumerating various interpretive options open to courts); Kosseff, *supra* note 43, at 779–85 (discussing moves made by some courts to narrow Section 230 immunity by treating online entities as coauthors of third-party content or as ineligible for protection against causes of action unrelated to publication).

[59] *See generally* Shlomo Klapper, *Reading Section 230*, 70 BUFF. L. REV. 1237 (2022).

[60] *See supra* Part II; *see also generally* Julio Sharp-Wasserman, Note, *Section 230(c)(1) of the Communications Decency Act and the Common Law of Defamation: A Convergence Thesis*, 20 COLUM. SCI. & TECH. L. REV. 195 (2018) (observing the similarity between Section 230 immunity doctrine and the pre-internet defamation defenses and speculating that the law of defamation would have produced similar outcomes even without Section 230).

[61] *See* Biden v. Knight First Amendment Inst. at Columbia Univ., 141 S. Ct. 1220, 1221–27 (2021) (Thomas, J., concurring).

That odd result is attributable to a deep tension between paragraphs (c)(1) and (c)(2) of the provision. Section 230(c)(2) expressly immunizes entities from liability on account of "any action voluntarily taken in good faith to restrict access to or availability of [objectionable material]." [65] Standing alone, the provision would operate as intended. Online entities would be free to remove offensive content with no fear of incurring legal liability, for paragraph (c)(2) provides a safe haven for the "'Good Samaritan' blocking and screening of offensive material."[66] But in practice entities rarely rely on (c)(2)'s protection of good-faith content moderation because courts have interpreted paragraph (c)(1) capaciously to protect online entities against any cause of action related to third-party-created content—regardless whether they screen for offensive material.[67] In practice, Section 230 makes online entities "indifferent to the content of information they host or transmit," for "whether they do (subsection (c)(2) or do not (subsection (c)(1)) [moderate content] there is no liability under either state or federal law."[68]

---

[62] *See* Chicago Lawyers' Comm. for C.R. Under L., Inc. v. Craigslist, Inc., 519 F.3d 666 (7th Cir. 2008) (Easterbrook, C.J.) (observing that, rather than encouraging content moderation, the dominant interpretation of Section 230(c)(1) makes entities "indifferent to the content of information they host or transmit") (quoting Doe v. GTE Corp., 347 F.3d 655 (7th Cir. 2003)); Zeran v. Am. Online, Inc., 129 F.3d 327 (4th Cir. 1997) (interpreting Section 230(c)(1) to operate as an independent ground for immunity, without reference to Section 230(c)(2)'s protection for good-faith moderation); *see also* Klapper, *supra* note 59 at 1252–56 (tracing this view to the *Zeran* decision's purposive approach to interpretation).
[63] *See supra* Part III.A.
[64] *See* Kosseff, *supra* note 2, at 61–64 (explaining that Section 230 was an alternative to Exon's bill because it addressed "the problem of online indecency with a different solution").
[65] 47 U.S.C. § 230(c)(2).
[66] 47 U.S.C. § 230(c).
[67] *See* Barnes v. Yahoo!, Inc., 570 F.3d 1096, 1105 (9th Cir. 2009) (observing that, as interpreted by the courts, "Subsection (c)(1), by itself, shields from liability all publication decisions").
[68] Doe v. GTE Corp., 347 F.3d 655, 660 (7th Cir. 2003) (Easterbrook, J.).

That is not an impossible reading of the statute. Section 230 (c)(1) flatly states that an online entity "shall not be treated as the publisher or speaker of any information provided by another."[69] Courts were presented with a difficult puzzle when asked to reconcile that text with (c)(2)'s more limited protection only for entities that moderate content. The point is that courts could easily have decided the issue differently. A plausible interpretation, perhaps more plausible, would read (c)(2), rather than (c)(1), as Section 230's key provision, immunizing only "Good Samaritans" that moderate offensive content, while reading (c)(1) merely to restate the pre-internet principle that an online entity shall not "be treated as the publish or speaker" of that information when it merely provides access to "information provided by another.".[70]

### 2. Immunity for Knowing and Intentional Wrongs

Second, in a significant departure from pre-internet mass-media law,[71] courts have interpreted Section 230 to allow online entities to claim immunity from legal liability even if they possess actual knowledge of unlawful material on their systems and still fail to remove it.[72] This may sound strange, but it, also, is not an impossible reading of the statute. Section 230(c)(1)'s sparse text, after all, includes no express limitation on the defendant's mental state. Further, although the paragraph's effect *is* limited to claims involving "publication" (and, most likely, should have been interpreted to refer to the common-law distinction between publishers on the one hand and distributors and conduits on the other, therefore applying only to publishers), courts have interpreted the paragraph to refer to any claim premised on an act of publication. Thus read, claims against distributors and conduits also fall within paragraph (c)(1)'s ambit, for those claims, too, must involve some ultimate element of publication.[73]

---

[69] 47 U.S.C. § 230(c)(1).
[70] *See* Klapper, *supra* note 59, at 1263–74 (2022) (examining the legislative materials and providing an extensive textualist defense of this interpretation). This view also accords with the original structure of the Cox-Wyden proposal, in which the text that now forms paragraph (c)(1) was placed immediately following the title of subsection (c), "Protection for 'Good Samaritan' Blocking and Screening of Offensive Material," rather than as a separate paragraph. *See* Internet Freedom and Family Empowerment Act, H.R. 1978, 104th Cong. (1995).
[71] *See supra* Part II.
[72] The leading case is Zeran v. Am. Online, Inc., 129 F.3d 327 (4th Cir. 1997), which interpreted Section 230 to immunize AOL from defamation liability despite its failure, even after notice, to remove defamatory messages from its system.

Probably more important to courts, this interpretation also supported the early internet's free-information ethos, which called for maximal freedom of expression online. Courts feared what is known as the heckler's veto problem: If platforms become liable for any content they are made aware of but fail to take down, platforms might decide to automatically take down, without investigation, any content even reported to them as objectionable to avoid the cost of investigating.[74] An internet user's post might be taken down and her freedom to speak her mind undermined by the unverified complaint of an internet "heckler." To avoid this problem and thereby further a policy of "freedom of speech in the new and burgeoning Internet medium," early courts granted broad immunity even where an entity is made aware of unlawful content and chooses not to remove it.[75]

### 3. Application to Nonpublication Claims

Third, courts have interpreted Section 230 to apply to nonpublication claims. Although Section 230 is publication centric—it encourages censorship and speaks in terms of "publisher or speaker" and "content provider"—publication is far from the modern internet's only or even primary function.[76] The internet operates as a complete virtual world, which includes not only publication-related wrongs, like defamation, but also physical-world wrongs, like designing defective smartphone apps or facilitating sex trafficking or illegal gun sales. Rather than argue that an online entity should have reviewed and moderated third-party content, such claims are analogous to physical-world product defect or conspiracy cases: They argue that an online entity should have designed its app or website differently, typically to include more safety features, or that it intentionally facilitated and profited from unlawful activity.[77] Courts considering such claims, however, sometimes ignore the distinction and grant Section 230 immunity defenses even for claims that do not allege a failure to review third-party content and thus do not implicate the content moderation burden and heckler's veto concern that inspired courts' broad interpretation of Section 230 in the first place.[78]

---

[73] *See Zeran*, 129 F.3d at 331 (so reasoning and citing William Prosser's classic treatise, which notes that publication is an element of claims both against distributors and against publishers); W. PAGE KEETON ET AL., PROSSER AND KEETON ON THE LAW OF TORTS § 113 (5th ed. 1984).
[74] *See generally* Brett G. Johnson, *The Heckler's Veto: Using First Amendment Theory and Jurisprudence to Understand Current Audience Reactions Against Controversial Speech*, 21 COMM. L. & POL'Y 175 (2016) (discussing the concept of the heckler's veto, whereby an individual is able to restrict another's freedom to speak by filing complaints against, shouting down, heckling, threatening, or otherwise harassing the speaker); *see also* Reno v. ACLU, 521 U.S. 844, 880 (1997) (invalidating portions of the Communications Decency Act, because, among other reasons, the requirement not to communicate indecent speech to "specific persons" "would confer broad powers of censorship, in the form of a 'heckler's veto,' upon any opponent of indecent speech"); Rory Lancman, *Protecting Speech from Private Abridgement: Introducing the Tort of Suppression*, 25 SW. U. L. REV. 223, 253–55 (1996) (discussing the origin of the "heckler's veto" concept).
[75] *Zeran*, 129 F.3d at 331.
[76] 47 U.S.C. § 230(c)(1).

*    *    *

Internet immunity doctrine as crafted by the courts thus extends a good deal further than would be required by Section 230's text. Online entities are eligible for protection without making any effort to censor objectionable content, even where they have actual knowledge of unlawful content on their systems, and even against causes of action with little connection to the publication of content. Such expansive protections worked well in the internet's early years, but they have become increasingly controversial as the internet has grown to take on a wider role in society.

## IV.  SECTION 230 REFORM

### A.  Hybrid Wrongs

One important question for Section 230 internet-immunity doctrine is what to do in cases of concerted action between online entities and their users. The best example is the 2016 decision by the U.S. Court of Appeals for the First Circuit, *Doe v. Backpage.com, LLC*.[79] The case involved three underage girls who had become victims of sex trafficking and were forced into sex work. Their traffickers listed their services for sale on the website Backpage.com, which, the plaintiffs alleged, intentionally aided the sex traffickers in perpetrating and concealing their wrongdoing in order to take a cut of the profits.[80]

---

[77] *See, e.g.*, Force v. Facebook, Inc., 934 F.3d 53, 57–59 (2d Cir. 2019); L.W. through Doe v. Snap Inc., 675 F. Supp. 3d 1087, 1098–110 (S.D. Cal. 2023)
[78] *See* Dickinson, *Rebooting Internet Immunity*, *supra* note 1, at 372–81.
[79] 817 F.3d 12 (1st Cir. 2016).

According to the complaint, when Backpage's competitor, Craigslist, closed its "Adult Services" section in 2010 due to sex trafficking concerns, Backpage intentionally enhanced the "Escorts" section of its website to maximize its profits by making sex trafficking easier.[81] The plaintiffs alleged that Backpage had "deliberate[ly] structure[ed] . . . its website to facilitate sex trafficking" and that it had "tailored its posting requirements" and established "rules and processes governing the content of advertisements" in a way that encouraged and facilitated sex trafficking.[82] For example, Backpage removed postings made as part of law enforcement sting operations and removed metadata from escort photographs to limit their usefulness to law enforcement agencies.[83] The plaintiffs alleged that Backpage made such moves to profit from sex traffickers' use of the website.[84] The "Adult" section was the only section of Backpage's site that charged a posting fee.[85] And for an extra fee, Backpage allowed users to post "Sponsored Ads" that appeared on the right hand side of every page in the "Escorts" section and included a picture of the advertised individual as well as her location and availability.[86]

The underage girls who brought suit asserted what was essentially a civil conspiracy claim against Backpage under the Trafficking Victims Protection Act,[87] which includes a private right of action against anyone who "knowingly benefits . . . from participation in a venture which that person knew or should have known has engaged in an act [of sex trafficking]."[88] The plaintiffs alleged that Backpage intentionally structured its website to aid the sex trafficking operations that had victimized them and to thereby maximize Backpage's

---

[80] *Id.* at 16.
[81] *See* Second Amended Complaint at ¶¶ 29–51, Doe v. Backpage.com, LLC, 104 F. Supp. 3d 149 (D. Mass. 2015) (No.14-13870) [hereinafter Backpage Complaint].
[82] *Backpage.com*, 817 F.3d at 16–17; *see* Backpage Complaint, *supra* note 81, at ¶¶ 52–59.
[83] *Backpage.com*, 817 F.3d at 16; Backpage Complaint, *supra* note 81, at ¶¶ 40, 51.
[84] Backpage Complaint, *supra* note 81, at ¶ 45.
[85] *Backpage.com*, 817 F.3d at 17; Backpage Complaint, *supra* note 81, at ¶ 43.
[86] *Backpage.com*, 817 F.3d at 17; Backpage Complaint, *supra* note 81, at ¶¶ 53–59.
[87] Victims of Trafficking and Violence Protection Act of 2000, Pub. L. No. 106–386, 114 Stat. 1464, div. A (2000) (codified as amended in scattered sections of the United States Code).
[88] 18 U.S.C. § 1595(a).

profits—the vast majority of which came from the "Escorts" section of its webpage.[89]

Although the case presented what, in the physical world, would be difficult questions about Backpage's alleged level of assistance to the sex traffickers and whether it knowingly participated in or benefited from a joint venture, the case was resolved straightforwardly under Section 230. The First Circuit concluded that Backpage was immune from civil liability because the claims related to the sex trafficker's escort listings on Backpage.com and therefore involved "information provided by [someone else]:" [90] "Whatever Backpage's motivations, those motivations do not alter the fact that the complaint premises liability on the decisions that Backpage is making as a publisher with respect to third-party content."[91]

The First Circuit's *Backpage.com* decision is no longer good law. In the wake of the decision, Congress enacted the Allow States and Victims to Fight Online Sex Trafficking Act of 2017 (FOSTA) to ensure that Section 230 would not apply to such lawsuits in the future.[92] Among other things, FOSTA authorized private actions against websites for facilitating sex trafficking by "publishing information designed to facilitate sex trafficking"[93] and prevented websites from invoking section 230(c)(1) to escape liability by adding subsection (e)(5), which specifically excludes violators from protection.[94]

However, the broader difficulties that the *Backpage.com* decision revealed in internet-immunity doctrine still remain. Because Section 230 was designed for a world of passive content intermediaries,[95] it does not consider an online entity's knowledge or intent.[96] Passive intermediaries merely convey the content of others, and holding them responsible for that content could impose a crippling content-moderation burden and spur them to censor user content. Thus, under Section 230, if the entity is not an author of content, it is immune from liability for harms that result from that content.[97] But online entities often do far more than act as passive intermediaries. Because Section 230's trigger for withholding immunity is content authorship, online entities can act in concert with their users in ways that cause harms, despite knowledge that those harms will occur or even intent to cause them, so long as they are not the authors of the content in question.

---

[89] *See Backpage.com*, 817 F.3d at 16.
[90] 47 U.S.C. § 230(c).
[91] *Backpage.com*, 817 F.3d at 21.
[92] Pub. L. No. 115-164, 132 Stat. 1253 (2018) (codified as amended in scattered sections of 18 U.S.C. & 47 U.S.C.).
[93] FOSTA § 5.
[94] *See* 47 U.S.C. § 230(e)(5).
[95] *See supra* Part III.B.
[96] *See* 47 U.S.C. § 230(c)(1); *see also, e.g.*, Daniel v. Armslist, LLC, 926 N.W.2d 710 (Wis. 2019) (explaining that plaintiff's allegation that Armslist knew its website was used for illegal gun sales "does not change the result" because section 230 "contains no good faith requirement" and "courts do not allow allegations of intent or knowledge to defeat a motion to dismiss").

The issue has been raised most squarely in a series of recent cases involving algorithmic curation and amplification of user-created content on social media. The Second Circuit's decision in *Force v. Facebook, Inc.* is a good example.[98] In that case, the surviving family members of victims of terrorist attacks in Israel brought suit against Facebook under the Antiterrorism Act (ATA),[99] which provides a private right of action against entities that, among other things, aid and abet acts of international terrorism[100] or provide material support[101] to terrorists or terrorist organizations.[102] The plaintiffs alleged that Facebook provided "material support" to the Hamas terrorist organization, in violation of the ATA, through its friend-suggestion algorithms, which helped connect prospective terrorists with others of similar ideology, and through its newsfeed algorithms, which had detected the terroristic tendencies of the victims' attackers and directed terrorism-related content toward them.[103]

Ordinarily, the viability of an aiding and abetting-type ATA claim like the *Force* plaintiffs would depend on whether helping to connect potential terrorists with Hamas leaders and providing them with links to radical literature constitutes sufficient "material support" to fall within the statute's prohibition.[104] Because of Section 230's immunity provision, however,

---

[97] *See* 47 U.S.C. § 230(c)(1).
[98] 934 F.3d 53 (2d Cir. 2019).
[99] Omnibus Diplomatic Security and Antiterrorism Act of 1986, Pub. L. No. 99–399, 100 Stat. 853 (1986); Antiterrorism and Effective Death Penalty Act of 1996, Pub. L. No. 104–132, 110 Stat. 1214 (1996) (relevant provisions of both codified as amended at 18 U.S.C. §§ 2331–2339D).
[100] 18 U.S.C. § 2333(d)(2).
[101] 18 U.S.C. §§ 2339A(a), 2339B(a)(1).
[102] *See Force*, 934 F.3d at 61.
[103] *Id.* at 59–59.
[104] *See, e.g.*, Benitez v. U.S. Att'y Gen., 543 F. App'x 913, 916 (11th Cir. 2013) (reasoning that

Facebook's key defense was not that it provided no material support, but that the plaintiff's claim treated it as a publisher or speaker of third-party-created content and that the operation of its own matching and newsfeed algorithms did not constitute content authorship.[105] The court resolved both questions in Facebook's favor, concluding first that the plaintiff's terrorism claims implicated Facebook's editorial discretion and thus treated it as a publisher or speaker of third-party-created content[106] and second that its use of algorithms to curate content and match users did not amount to content authorship.[107]

Like *Backpage*, the Second Circuit's decision in *Force* reveals Section 230's poor fit with the modern internet. Unlawful online conduct may relate to third-party-created content, but it may still be at least as analogous to traditional physical-world wrongdoing as it is to the electronic publication wrongs (like defamation) that Section 230 was designed for.[108] Dissenting in *Force*, Chief Justice Katzmann made a similar point. Suppose an acquaintance called you on the telephone and said "I've been reading over everything you've ever published . . . [and have] done the same for this other author" and "think you'd get along," so "here is a link to all her published works."[109] You "might say your acquaintance fancies himself a matchmaker," but you would not say your acquaintance is "acting as the *publisher* of the authors' work."[110] That under Section 230 the entire question of immunity turns on publication and authorship, even though much wrongdoing on the modern internet does not, has pressed courts into linguistic knots as they try to make the test work in contexts it was never designed for.[111]

---

"[t]he plain language of the material support bar lists 'transportation' as an example of material support, and [the appellant] provided the [Revolutionary Armed Forces of Colombia] with air transportation").

[105] *Force*., 934 F.3d at 62.
[106] *Id.* at 66–67.
[107] *Id.* at 68–71.
[108] *See* Dickinson, *Rebooting Internet Immunity*, *supra* note 1, at 372–81.
[109] *Force*, 934 F.3d at 76 (Katzman, J., concurring).
[110] *Id.*
[111] *See* Kosseff, *supra* note 43, at 779–85.

### B. *Legislative Reform Efforts*

Troubled by online wrongdoing that is not easily resolved by Section 230's content-authorship test for immunity, legislators have proposed dozens of statutory reforms to the aging statute. Many, like FOSTA, are targeted at misconduct that seems to be slipping through the cracks. Not long after the *Force v. Facebook* decision, for example, Representative Tom Malinowski introduced the Protecting Americans from Dangerous Algorithms Act, which would have amended Section 230 to remove protection against claims involving content amplification leading to civil rights violations or acts of international terrorism.[112]

Other proposals abound. For example, a Department of Justice review of Section 230 concluded that a "Bad Samaritan" carve out should be added to the statute to ensure that online entities that purposefully solicit "third parties to sell illegal drugs to minors, exchange child sexual abuse material," or engage in other unlawful activities through their platforms "do not benefit from Section 230's sweeping immunity at the expense of their victims."[113] A host of proposed amendments would strip immunity from online entities that facilitate such behavior. Academic commentators have also proposed numerous suggestions to address hybrid publication-nonpublication misconduct, such as limiting immunity to claims that impose a content-moderation burden on defendants, applying joint enterprise liability theory to allow claims against entities that intentionally profit from their users' misconduct, and imposing a reasonableness requirement that would require online entities to take reasonable steps to prevent misuse of their services to qualify for immunity.[114] Such proposals are motivated by the perceived overprotection of online entities under Section 230, particularly for the sorts of hybrid publication-nonpublication wrongs discussed above.

Other proposals, rather than amend Section 230 to fill perceived gaps, would leverage the statute's importance to technology companies to press them to take action on other issues. For example, in 2024, Senators Hawley and

---

[112] H.R. 2154, 117th Cong. (2021).
[113] *See* U.S. DEP'T OF JUST., SECTION 230—NURTURING INNOVATION OR FOSTERING UNACCOUNTABILITY? 3, 14–15 (2020), https://perma.cc/68FP-5UUC.
[114] *See* Dickinson, *The Internet Immunity Escape Hatch*, *supra* note 1, at 1448–50 (collecting and discussing various reform proposals).

Blumenthal introduced the No Section 230 Immunity for AI Act, which would revoke Section 230 online-intermediary immunity for claims predicated on an online entity's use of "generative artificial intelligence tools."[115] By selectively eliminating a powerful defense, the bill seeks to deter the misuse of generative AI tools like ChatGPT and DALL-E to create "deepfakes" to deceive consumers or voters.[116] Similarly, the Biased Algorithm Deterrence Act would target platforms' promotion and demotion of select user content by withholding Section 230 immunity for claims against online entities stemming from user-created content that the entity reordered to appear other than as chronologically posted.[117] The variations are countless, but all follow the same basic pattern: Press powerful technology companies to take action on preferred issues by threatening to eliminate a valuable litigation defense.

Naturally, some reform proposals are much more promising than others. Small amendments, like FOSTA's sex trafficking amendment, may correct for particularly egregious disparities between online and physical-world vicarious liability regimes; but they are not a long-term solution.[118] Section 230's shortcomings extend to all manner of hybrid publication-nonpublication wrongs, not any single type. Eventually, broad reform will be required that updates Section 230 for the modern, non-publication-centric internet. In the meantime, legislatures would do well to reject the innumerable proposals to amend Section 230 to pressure tech companies into action on unrelated policy issues, where the prospect of vicarious liability has not traditionally served as a check on the disfavored behavior. For example, proposals to remove liability protection for the transmission of content authored by the Chinese Communist

---

[115] A Bill to Waive Immunity Under Section 230 of the Communications Act of 1934 for Claims and Charges Related to Generative Artificial Intelligence ("No Section 230 Immunity for AI Act"), S. 1993, 118th Cong. (2023).

[116] *See* Katie Paul, *Bipartisan U.S. Bill Would End Section 230 Immunity for Generative AI*, REUTERS, June 14, 2023, https://perma.cc/K5GD-QYXJ.

[117] H.R. 492, 116th Cong. (2020). For criticism of such technology-focused statutes as likely to overdeter socially beneficial economic activity, *see* Gregory M. Dickinson, *The Patterns of Digital Deception*, 65 B.C. L. REV. 2457, 2480–97 (2024).

[118] Whether criminalization of sex work and Congress's amendment of Section 230 to permit civil claims truly benefits sex workers is a separate question. Criminalizing sex work and purging related content from the internet may harm sex workers by forcing them to work in secret, more dangerous environments. *See* Citron, *How to Fix Section 230*, *supra* note 14, at 736–42.

Party [119] or other foreign adversaries, [120] for censorship of constitutionally protected speech,[121] or for platform action based on a user's political affiliation or other characteristics[122] do not aim to correct imbalances in the vicarious liability rules for online versus offline conduct, but to extract concessions from the tech industry on partisan political issues.[123] Threats to poke holes in Section 230 immunity may give legislators leverage over tech companies, but such amendments are unlikely to produce a coherent vicarious liability regime.[124] Either Section 230 immunity is good policy for internet liability, or it is not. If it is good policy, it should be generally available to all internet companies; if it is bad policy, it should be eliminated.

Concern over ill-advised amendments may be more theoretical than practical, however, for widespread disagreement on internet policy and stalemate in Congress mean that any legislative reform faces an uphill battle.[125]

---

[119] *See* Act to Amend Section 230 of the Communications Act of 1934 to Exclude from the Application of Such Section Persons or Entities that are Owned by the Chinese Communist Party, H.R. __, 117th Cong. (discussion draft of the H. Comm. on Energy & Com., 2021), https://perma.cc/2WDA-52ZL.

[120] *See* Removing Section 230 Immunity for Official Accounts of Censoring Foreign Adversaries Act, S. 941, 118th Cong. (2023).

[121] *See* Letter from Eugene Volokh, Professor of Law, University of California, Los Angeles, to the House Subcommittee on Communications and Technology (Nov. 30, 2021) (letter commenting on draft Preserving Constitutionally Protected Speech Act, H.R. __, 117th Cong. (2021)), https://perma.cc/H68F-P34J.

[122] *See* Act to Amend Section 230 of the Communications Act of 1934 to Limit Immunity under Such Section for Actions Based on Racial, Sexual, Political Affiliation, or Ethnic Grounds, H.R. __, 117th Cong. (discussion draft of the H. Comm. on Energy & Com., 2021), https://perma.cc/EF3Y-8LH7.

[123] For a lengthy collection of proposed bills to amend Section 230, *see Section 230 Legislation Tracker*, LAWFARE, https://perma.cc/CA5J-HQG8.

[124] *See generally* Fred S. McChesney, *Rent Extraction and Rent Creation in the Economic Theory of Regulation*, 16 J. LEGAL STUD. 101 (1987) (discussing incentives for legislators to engage in "rent creation," that is, the extraction of concessions from industry participants through threats to impose costs).

[125] *See* Dickinson, *The Internet Immunity Escape Hatch*, *supra* note 1, at 1447–57 (describing various reform proposals and impediments to congressional action); *see also* Mary Clare Jalonick, *From Data Privacy to AI, Here are New Rules Congress is Considering for Tech Companies*, PBS (May 8, 2023, 3:16 PM), https://perma.cc/FYE4-38CY (noting that "[m]ost Democrats and Republicans agree that the federal government should better regulate the biggest technology companies," but that "there is very little consensus on how it should be done"); Zach Schonfeld & Rebecca Klar, *Supreme Court Punts Section 230 Debate Back to Congress*, THE HILL (May 22, 2023, 3:18 PM), https://perma.cc/FPB9-2K5R (explaining that "despite bipartisan criticism" of Section 230 "lawmakers face a stalemate on how to reform it").

Although there is broad support for reform across the political spectrum, Democrats and Republicans focus on very different concerns. Democrats criticize online platforms' failure to protect the public from harmful content such as vaccine misinformation, falsified political ads, hate speech, and materials promoting terrorism,[126] whereas Republicans focus on perceived political bias, alleging that platforms' censorship practices disproportionately affect conservative voices.[127] Moreover, even if Congress were to agree on particular policy goals, it is notoriously difficult to craft effective legislation to govern rapidly evolving technology.[128] Even when it has the time and political support to do so, members of Congress often lack the expertise [129] to understand and regulate complex technological issues and are likely to be

---

[126] *See* Steven Lee Myers, *Case Will Test Effort to Fight Misinformation*, N.Y. TIMES, Feb. 10, 2023, at A1 (discussing Biden administration's efforts to limit spread of misinformation regarding COVID-19); Tony Romm & Elizabeth Dwoskin, *Silicon Valley Braces for Regulatory Change*, WASH. POST, Jan. 19, 2021, at A1 (discussing Democrats' and President Biden's efforts to press online platforms to censor online content).

[127] *See* Will Oremus, The Messy Battle Over Online Speech, WASH. POST, Oct. 30, 2022, at G1 (discussing Republican criticism of platform censorship and Elon Musk's decision to purchase Twitter)*;* Tony Romm, *GOP Votes to Subpoena Facebook, Twitter CEOs*, WASH. POST, Oct. 23, 2020, at A16 (discussing the Republicans' "campaign to rethink Section 230" "in response to concerns about political bias").

[128] *See generally* Jody Freeman & David B. Spence, *Old Statutes, New Problems*, 163 U. PA. L. REV. 1 (2014); Wulf A. Kaal & Robert N. Farris, *Innovation and Legislation: The Changing Relationship—Evidence from 1984 to 2015*, 58 JURIMETRICS J. 303, 304 (2018); Albert C. Lin, *Technology Assessment 2.0: Revamping Our Approach to Emerging Technologies*, 76 BROOK. L. REV. 1309 (2011).

[129] Recognizing its own shortcoming, Congress in 1972 established the Office of Technology Assessment (OTA) to "equip [the legislative branch] with new and effective means for securing competent, unbiased information concerning the physical, biological, economic, social, and political effects of [technology]" and to use such information to provide "legislative assessment of matters pending before the Congress." Technology Assessment Act of 1972, Pub. L. No. 92-484, § 2, 86 Stat. 797 (1972) (codified at 2 U.S.C. §§ 471–81). Congress abolished the OTA in 1995 by cutting off its funding. It may be time to reconsider that decision. *See* Kevin Kosar, *Congress's Tech Policy Knowledge Gap*, CATO UNBOUND (June 10, 2019), https://perma.cc/2FTV-LLNS (discussing the OTA's purpose and advocating for its revival); Patrick Healy & Cornelia Dean, *Clinton Says She Would Shield Science from Politics*, N.Y. TIMES, Oct. 5, 2007, at A22 (reporting then-Senator and Democratic presidential candidate Hillary Clinton's call in 2007 for Congress to revive the OTA).

heavily influenced by industry lobbyists and other beneficiaries of the status quo.[130]

### C. The Judicial Path

Even without action from Congress, change is in the air. The Supreme Court has had numerous opportunities in recent years to offer its interpretation, but it has not yet stepped into the discussion. That may soon change. In 2020, after decades of the Court declining to hear cases regarding Section 230, Justice Thomas took the unusual step in *Malwarebytes v. Enigma* of issuing a statement to explain why, "in an appropriate case," the Supreme Court should take up a Section 230 case to consider the appropriate scope of internet immunity.[131] He explained that "[c]ourts have long emphasized nontextual arguments when interpreting §230, leaving questionable precedent in their wake."[132] In particular, he questioned courts' application of Section 230 immunity even to the sorts of hybrid publication-nonpublication contexts discussed above: to platforms that leave content on their sites that they know to be unlawful; to those that algorithmically curate, seek out, or promote unlawful material; and to claims outside the publishing context, such as those related to defective products. Sensing a gap between Congress's words and current internet immunity doctrine, Justice Thomas urged the Court in a future case to consider whether "the text of [Section 230] aligns with the current state of immunity enjoyed by Internet platforms."[133]

Section 230 was poised for long-awaited change to materialize when in 2022 the Supreme Court agreed to review *Gonzalez v. Google LLC*[134] and *Twitter, Inc. v. Taamneh*.[135] Both cases involved acts of international terrorism, one in Paris and the other in Istanbul, for which surviving family members sought to hold online platforms accountable. The plaintiffs alleged that the platforms aided and abetted and provided material support to terrorists in violation of the ATA through their content-curation algorithms, which allowed ISIS to post recruitment videos on the platforms and promoted that content to individuals most likely to be interested in its message.[136] As in *Force v. Facebook*, the defendants argued that they were immune from suit under Section 230 because they had not authored the content in question, merely recommended and conveyed it to others.[137]

---

[130] *See generally* George J. Stigler, *The Theory of Economic Regulation*, 2 BELL J. ECON. & MGMT. SCI. 3 (1971); Sam Peltzman, *Toward a More General Theory of Regulation*, 19 J.L. & ECON. 211 (1976); MANCUR OLSON, THE LOGIC OF COLLECTIVE ACTION: PUBLIC GOODS AND THE THEORY OF GROUPS (1965).
[131] Malwarebytes, Inc. v. Enigma Software Grp. USA, LLC, 141 S. Ct. 13, 14 (2020)..
[132] *Id.*
[133] *Id.*
[134] 598 U.S. 617 (2023).
[135] 598 U.S. 471 (2023).

Despite expectations for a momentous change in internet immunity doctrine, the Supreme Court resolved both cases without addressing the scope of Section 230 at all. In a pair of decisions issued in May 2023, the Court concluded that the plaintiffs failed to allege plausible theories of relief under the ATA and therefore the Court had no need to consider the viability of the platforms' Section 230 defenses.[138] Internet policy experts gave a collective sigh of relief and disappointment—relief that the Supreme Court had not risked radical change to a key statute and disappointment that the wait for internet-immunity reform would continue to some future day.[139]

The wait is unlikely to be long. With so much human activity now conducted over the internet, the potential liability of online entities for facilitating online wrongdoing is and will remain a central concern. Especially important are outstanding questions such as the extent to which online entities can be held liable for their algorithmic amplification of unlawful material and whether they can face product liability claims for the design of apps and services that aggregate and monetize third-party-created content. Lawsuits around the country continue to pursue these and other theories of relief against online platforms, many of which are among the most prominent and powerful companies in the world.[140] Neither Congress nor the Court can put off the questions forever.

---

[136] *See Gonzalez*, 598 U.S. at 555; *Taamneh*, 598 U.S. at 474.
[137] *See* Brief of the Respondent Google LLC at 21–33, *Gonzalez*, 598 U.S. 617 (2023) (No. 21–1333); Brief of Petitioner Twitter, Inc. at 14–15, *Taamneh*, 598 U.S. 471 (2023) (No. 21-1496).
[138] *See Taamneh*, 598 U.S. at 506–07 (concluding plaintiffs failed to state a cognizable claim and reversing decision below); *see also Gonzalez*, 598 U.S. at 556 (vacating and remanding for consideration in light of *Taamneh* because the complaint "appears to state little, if any, plausible claim for relief").
[139] *See, e.g.*, Robert Barnes & Cat Zakrzewski, *High Court Sides with Big Tech in Content Suit*, WASH. POST, May 19, 2023, at A1 (contrasting tech companies' and free-speech advocates' satisfaction with the rulings with the disappointment of a nonprofit group advocating for change, which said the decision "rewards Big Tech for bad conduct"); Daphne Keller, *Stanford's Daphne Keller on SCOTUS Decision that Google, Twitter, and Facebook not Responsible for Islamic State Deadly Posts*, STAN. L. SCH. (May 19, 2023), https://perma.cc/9HJD-QG6Q ("In a way this is an 'everybody, calm down' moment. The court strongly affirmed that basic tort principles apply and protect platforms, just like they protect other communications services.").

## V. CONCLUSION

With Congress at a standstill, reform is most likely to come through the Supreme Court. Whenever it does finally take up the Section 230 question, it will be writing on a completely blank slate. Because it has never before interpreted the statute, all avenues are open to it. It could endorse the prevailing interpretation found in existing case law, flaws and all, perhaps on the ground that Congress, not the judiciary, should lead any reform. It could make small tweaks to immunity doctrine where it perceives past courts to have gone astray. Or, perhaps just as likely, it could upend the past cases and take internet immunity doctrine in an entirely new direction.

The Court will be pulled in two directions. On the one hand, the Court is acutely aware of its limitations regarding technology policy. As Justice Kagan put it during oral argument of *Gonzalez v. Google*, "We really don't know about these things. You know, [we] are not like the nine greatest experts on the internet."[141] The importance of the question and the Court's awareness of its own limitations counsel incremental rather than radical change. Furthermore, with broad internet immunity now well-entrenched in the state and federal courts, the Supreme Court may be hesitant to disrupt the current system.

Yet a close, textual reading of Section 230, the Court's preferred method, supports far more sweeping change. If the Court does opt for radical reform, ardent supporters of the status quo can take at least some comfort in the legal system's numerous other checks on intermediary liability. Section 230 is a defense, after all, not a right of action. Even were it completely eliminated, a prospective plaintiff could not succeed without pleading a plausible cause of action (which the plaintiffs in *Gonzalez* and *Taamneh* notably failed to do) and overcoming the conduit liability defense and other common-law limitations on intermediary liability that foreshadowed and now continue to serve as an anti-liability backstop to Section 230. Important as the statute may be, the internet will hold fast, and floodgates will not burst. The law is far too robust for that.

---

[140] *See* Isaiah Poritz, *Social Media Addiction Suits Take Aim at Big Tech's Legal Shield*, BLOOMBERG (Oct. 25, 2023, 1:14 PM), https://perma.cc/HBM9-MJ6W (discussing numerous lawsuits alleging social-media platforms have caused a youth mental health crisis and platforms' Section 230 defenses); *e.g.*, Doll v. Richard Pelphrey & Grindr Inc., No. 23 CI 00363, 2024 Ky. Cir. LEXIS 84 (Ky. Cir. Ct. Oct. 18, 2024); In re Soc. Media Adolescent Addiction/Pers. Inj. Prod. Liab. Litig., __ F. Supp. 3d __, No. 23-CV-05448, 2024 WL 4532937 (N.D. Cal. Oct. 15, 2024); Anderson v. TikTok, Inc., 116 F.4th 180 (3d Cir. 2024).

[141] Transcript of Oral Argument at 46, *Gonzalez*, 598 U.S. 617 (2023) (No. 21–1333).